\documentclass[aps,prd,floatfix,twocolumn,superscriptaddress,preprintnumbers,nofootinbib]{revtex4-1}

\usepackage{textcomp} 
\usepackage{slashed}
\usepackage{epsfig,latexsym,cancel,amssymb,amsmath,verbatim,mathrsfs}
\usepackage{color}
\usepackage{graphicx}
\usepackage{enumitem}
\usepackage[colorlinks,citecolor=blue]{hyperref}

\newcommand{\beq}{\begin{equation}}
\newcommand{\eeq}{\end{equation}}
\newcommand{\bea}{\begin{eqnarray}}
\newcommand{\eea}{\end{eqnarray}}

\allowdisplaybreaks[1]

\begin{document}

\title{Probing the $Hgg$ coupling through the jet charge correlation in Higgs boson decay}

\author{Xiao-Rui Wang}
\email{xiaorui\_wong@pku.edu.cn}
\affiliation{Department of Physics and State Key Laboratory of Nuclear Physics and Technology, Peking University, Beijing 100871, China}

\author{Bin Yan}
\email{yanbin@ihep.ac.cn (corresponding author)}
\affiliation{Institute of High Energy Physics, Chinese Academy of Sciences, Beijing 100049, China}

\begin{abstract}
The effective coupling of the Higgs boson to gluons determines the production rate of the Higgs boson at the LHC and plays a crucial role to measure the Higgs properties. In this paper, we propose a novel method to extract the Higgs gluon gauge coupling information at the future lepton colliders, by utilizing the jet charge asymmetry of the two leading jets from the Higgs boson decay.
Owing to the jet charge correlation nature of the Higgs decay products, we demonstrate that this asymmetry value would be close to one for the signal $H\to gg$, while it is always larger than one for the dominated backgrounds. With the help of the jet charge information, it is possible to constrain the Higgs gluon effective coupling very well in the future Higgs factory and this conclusion is not sensitive to the jet charge parameter $\kappa$.
\end{abstract}

\maketitle

\section{Introduction}
The discovery of the Higgs boson at the Large Hadron Collider (LHC)~\cite{ATLAS:2012yve,CMS:2012qbp} has opened a new era in particle physics. The precise measurements of the Higgs couplings play a crucial role to test the Standard Model (SM) and beyond~\cite{Cheung:2014noa,Englert:2014uua,Cao:2015oaa,Falkowski:2015fla,Cao:2015oxx,Corbett:2015ksa,Cao:2016zob,Durieux:2017rsg,Cao:2018cms,DeBlas:2019qco,Li:2019uyy,Ellis:2020unq,Xie:2021xtl,Yan:2021tmw,Dong:2022bkd}. From the recent global analysis of the ATLAS and CMS Collaborations at the LHC Run-2~\cite{ATLAS:2022vkf,CMS:2022dwd}, it shows that the observed properties of the Higgs boson agree with the SM prediction very well. Owing to the dominant production channel of the Higgs boson at the LHC being the top quark loop-induced gluon-fusion process, the effective coupling of the Higgs boson to gluons has received much attention in the high energy physics community. It shows that the effective coupling strength modifier $\kappa_g$ of the $Hgg$ coupling has been strongly constrained by the Higgs signal strength measurements at the LHC Run-2~\cite{ATLAS:2022vkf,CMS:2022dwd}, {\it i.e.,} $\kappa_g=0.95\pm 0.07$ (ATLAS) and $\kappa_g=0.92\pm 0.08$ (CMS). Here the effective Lagrangian of the $Hgg$ coupling is parameterized as,
\beq
\mathcal{L}_{Hgg}=\kappa_g\frac{\alpha_s}{12\pi v}HG_{\mu\nu}^aG^{a\mu\nu},
\eeq
where $v=246~{\rm GeV}$ is the vacuum expectation value and $\alpha_s$ is the strong coupling. $G_{\mu\nu}^a$ is the field strength tensor of the gluon with the color index $a$ and $\kappa_g=1$ for the SM. 

However, the limit for the $Hgg$ coupling at the LHC is inferred from the signal strength measurement of the gluon-fusion Higgs production, which would strongly depend on the assumption of the Higgs decay information. A more straightforward strategy to probe the $Hgg$ coupling is measuring the signal strength of $H\to gg$. Unfortunately, it would be a challenging task at the LHC due to the overwhelmingly large QCD background. But the signature of the Higgs boson at the $e^+e^-$ collider is clean and readily identifiable, therefore, we could directly extract the $Hgg$ coupling information from the Higgs decay with very high accuracy. Currently,  several candidates for the lepton colliders have been proposed by the high energy physics community, such as the Circular Electron Positron Collider (CEPC)~\cite{CEPCStudyGroup:2018ghi}, the Future Circular Collider (FCC-ee)~\cite{FCC:2018evy}, and the International Linear Collider (ILC)~\cite{ILC:2013jhg}. With the help of the high tagging efficiencies  of the heavy quarks at the lepton collider, the uncertainty of the $\kappa_g$ could be reduced to the $\mathcal{O}$(1\%) level~\cite{An:2018dwb}. Although the Higgs boson production can be disentangled from its decay by the recoil mass method at the lepton collider, the measurements of the Higgs couplings are still depending on the assumption of the Higgs width. One of the approaches to avoid the impact of the Higgs width without the global fitting is to measure the jet energy profile from the Higgs boson decay, which utilizes the different features of the soft gluon radiations between the quark jet and gluon jet~\cite{Li:2018qiy,Li:2019ufu}. It shows that the jet energy profile would be sensitive to the fractions of quark and gluon final states, while the dependence of the production cross section and the Higgs width would be canceled from the definition.

Recently, we demonstrated that the jet charge asymmetry from the two leading jets of the production of the Higgs+2 jet is sensitive to the charge correlations between the jets and can be used to discriminate the different Higgs production mechanisms at the LHC~\cite{Li:2023tcr}. With a similar idea, we will apply this observable to measure the $Hgg$ coupling at the lepton collider due to the uncorrelated nature of the jet charges in $H\to gg$, while the jet charges are correlated for the main backgrounds. As to be shown below, this observable would be
also only sensitive to the fractions of the different channels, and the possible new physics effects in Higgs production and total width would be canceled.
The jet charge asymmetry from the Higgs boson decay can be defined as~\cite{Li:2023tcr},
\beq
\overline{A}_Q\equiv\frac{\langle \left|Q_J^1-Q_J^2\right |\rangle}{\langle \left|Q_J^1+Q_J^2\right |\rangle} \equiv \frac{\langle Q^{(-)} \rangle  }{\langle Q^{(+)} \rangle},
\label{eq:Asy}
\eeq 
where $\langle Q\rangle$ denotes the average value of the quantity $Q$ and $Q^{(\pm)}=|Q_J^1\pm Q_J^2|$, with $Q_J^{1,2}$ is the jet charge of the leading and subleading jets, respectively. Therefore, we could expect that this asymmetry would be close to one for the signal $H\to gg$, while it would be larger than one for the backgrounds, {\it e.g.,} $H\to b\bar{b}, H\to VV^*$ with $V=W, Z$.

\section{The Jet Charge}
The jet charge can be defined as the  $p_T$ weighted sum of the electric charge of the jet constituents~\cite{Field:1977fa,Krohn:2012fg,Waalewijn:2012sv},
\beq
Q_J=\frac{1}{(p_T^J)^\kappa}\sum_{i\in {\rm jet}}Q_i(p_T^i)^\kappa,\quad \kappa>0,
\eeq
where $p_T^J$ is the transverse momentum of the jet,  $p_T^i$ and $Q_i$ are the transverse momentum and electric charge of particle $i$ inside the jet. The parameter $\kappa>0$ is used to suppress the contribution from the soft gluon radiation.  One important feature of the jet charge is that it could serve well for identifying the charge of the primordial parton of the hard scattering, as result, this observable has been widely discussed in the literature from both the theoretical calculation~\cite{Krohn:2012fg,Waalewijn:2012sv} and experimental measurements~\cite{CMS:2014rsx,ATLAS:2015rlw,CMS:2017yer,CMS:2020plq}. It shows that the theoretical prediction of $Q_J$ at the next-to-leading order accuracy of QCD agrees with the measurements from the ATLAS and CMS Collaborations very well~\cite{ATLAS:2015rlw,CMS:2017yer}.  
Recently, this observable has been widely used for discriminating the quark jet and gluon jet ~\cite{Fraser:2018ieu,Larkoski:2019nwj,Gianelle:2022unu}, searching for the possible new physics effects~\cite{Chen:2019uar,Li:2021uww,Li:2023tcr}, probing nuclear  medium effects in heavy-ion and electron-ion collisions~\cite{Chen:2019gqo,Li:2019dre,Li:2020rqj}, as well as probing the quark flavor structure of the  nucleon~\cite{Kang:2020fka,Kang:2021ryr,Lee:2022kdn,Kang:2023ptt}. In this work, we will demonstrate below that the jet charge information could also be used to constrain the $Hgg$ effective coupling in Higgs boson decay and this information would be complementary to that obtained from the jet energy profile and cross section measurements at the lepton collider in determining the $Hgg$ coupling.

\section{Collider Simulation}
In this section, we perform a detailed Monte Carlo simulation to explore the potential of probing the $Hgg$ coupling through the jet charge asymmetry at the future CEPC with the center-of-mass energy $\sqrt{s}=250~{\rm GeV}$. At the CEPC, the Higgs boson is dominantly produced through the Higgs and $Z$ boson associated production, {\it i.e.,} $e^+e^-\to H Z$. The electroweak and QCD corrections for this process have been widely discussed in recently years~\cite{Gong:2016jys,Sun:2016bel,Chen:2018xau,Li:2020ign,Chen:2022mre}. The preliminary simulations have shown that the cross section of the $ZH$ production at the CEPC could be measured with an uncertainty of $0.51\%$ under the integrated luminosity of $\mathcal{L}=5.6~{\rm ab}^{-1}$~\cite{CEPCStudyGroup:2018ghi}. The production of the millions of the Higgs boson at the CEPC would offer the opportunity for precisely measuring the $Hgg$ coupling directly. The signal  we are interested in is 
Higgs decaying into gluons and $Z$ boson decaying into lepton pairs ($e^+e^-$ and $\mu^+\mu^-$). The major SM backgrounds are $H(\to b\bar{b},c\bar{c})Z$, $Zjj$ and  $H(\to VV^*\to 4j)Z$. However, the branching ratio of $gg$ mode (${\rm BR}(H\to g g)=8.6\%$) is much smaller than the $b\bar{b}$ channel (${\rm BR}(H\to b\bar{b})=58.2\%$)~\cite{Workman:2022ynf}, therefore, the flavor tagging technique would be necessary for suppressing the backgrounds from heavy quarks. It shows that the heavy flavor backgrounds can be removed mostly if we require two non-$b$ and $c$ jets in the final state with the identification efficiency of gluon jet $97.2\%$~\cite{Gao:2016jcm}.   
The background $Zjj$ can also be highly suppressed after we include the kinematic cuts, {\it e.g.,} recoil mass and the polar angle of the Higgs boson~\cite{Chen:2016zpw,CEPCStudyGroup:2018ghi}. As shown in Ref.~\cite{Gao:2016jcm}, after the kinematic cuts and requiring two light jets in the final state, we could get the signal events for the $H\to gg$ with $\kappa_g=1$ is $N_g=3438$ at $\sqrt{s}=250~{\rm GeV}$ with an integrated luminosity of $5.6~{\rm ab}^{-1}$.  The background of Higgs decaying into heavy flavor quarks ($H\to b\bar{b},c\bar{c}$) is about $N_{\rm HF}=0.1N_g$, the event number of $Zjj$ is $N_{\rm Zjj}=0.2N_g$, and the background of $H\to VV^*$ is $N_{\rm VV}=0.6~N_g$~~\cite{Gao:2016jcm}. Since the jet charge is a track-based observable, we expect that the jet charge asymmetry of we are considering would be not sensitive to the kinematic cuts in the above analysis and these event numbers will be used for rescaling the statistical error of the jet charge asymmetry and calculating the fractions of the different processes.

\begin{figure*}
\centering
\includegraphics[scale=0.2]{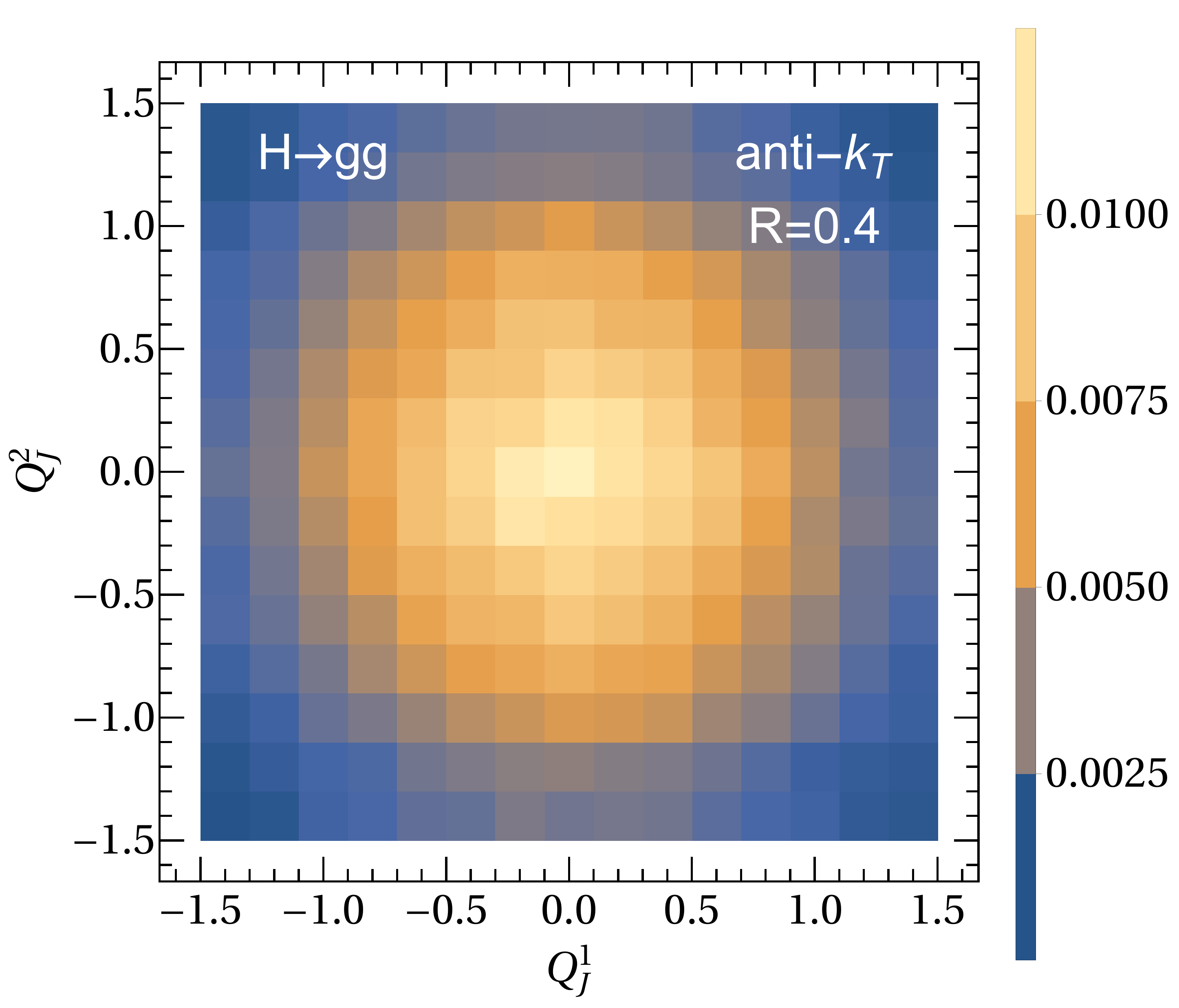}
\includegraphics[scale=0.2]{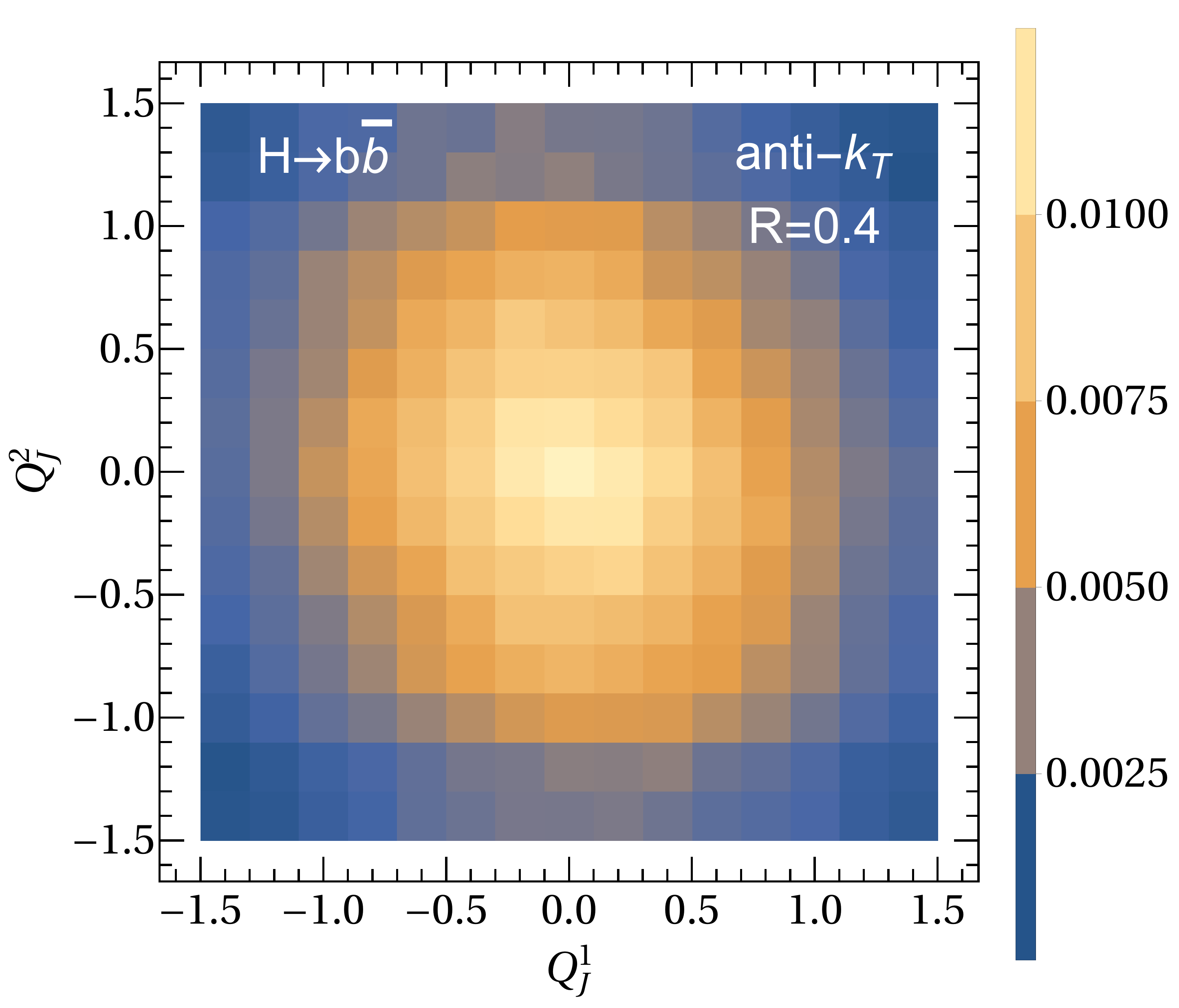}
\includegraphics[scale=0.2]{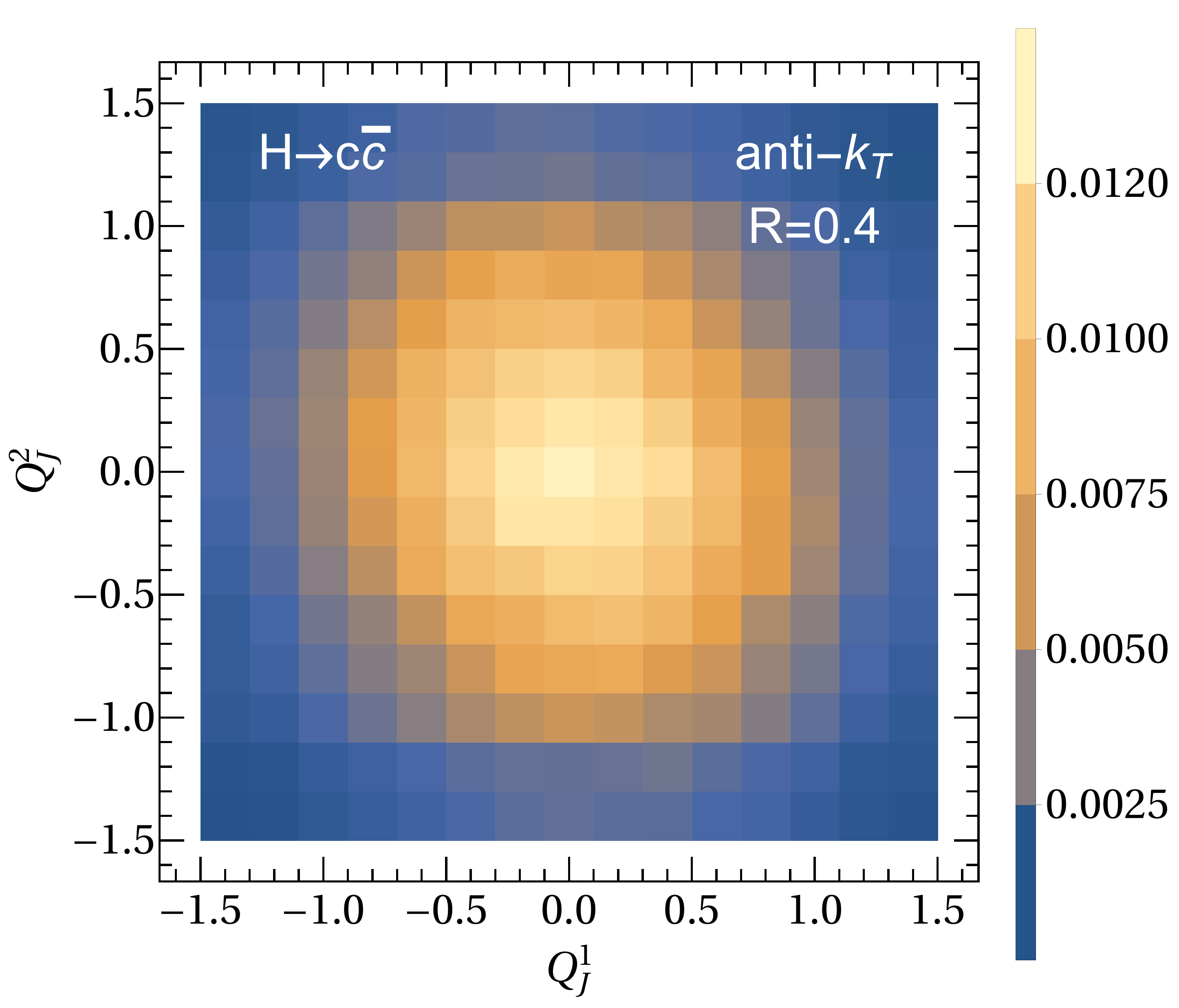}
\includegraphics[scale=0.2]{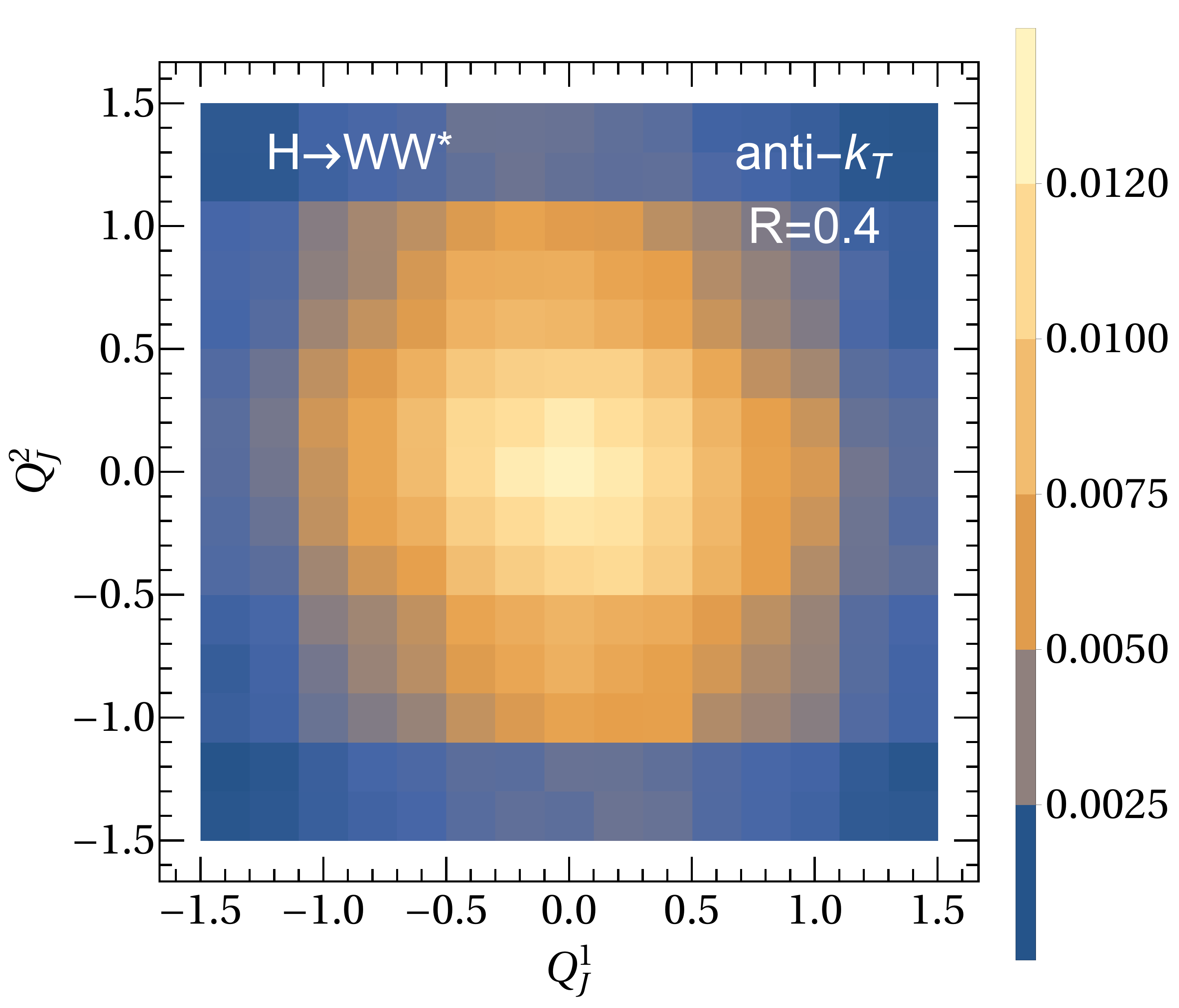}
\includegraphics[scale=0.2]{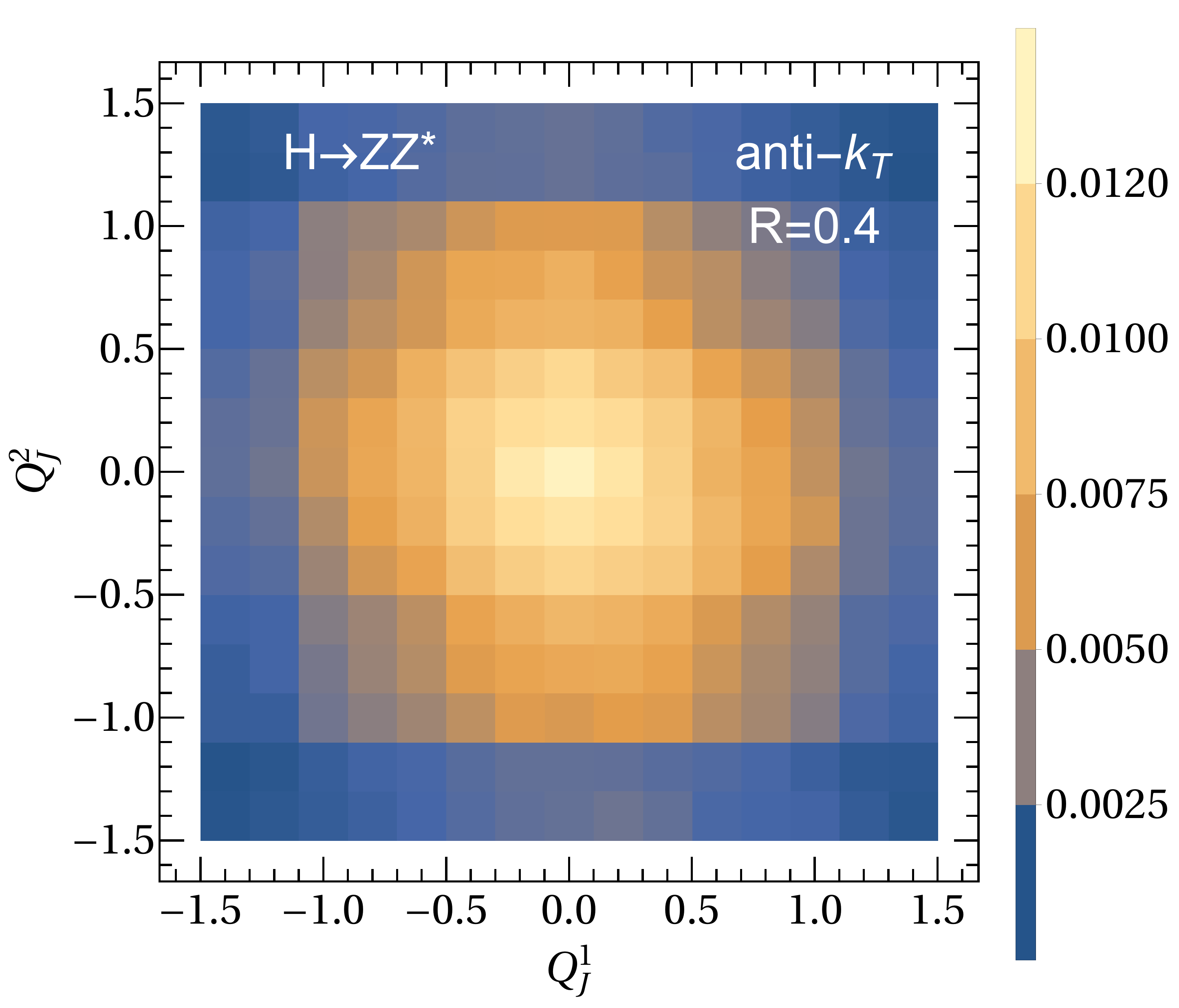}
\includegraphics[scale=0.2]{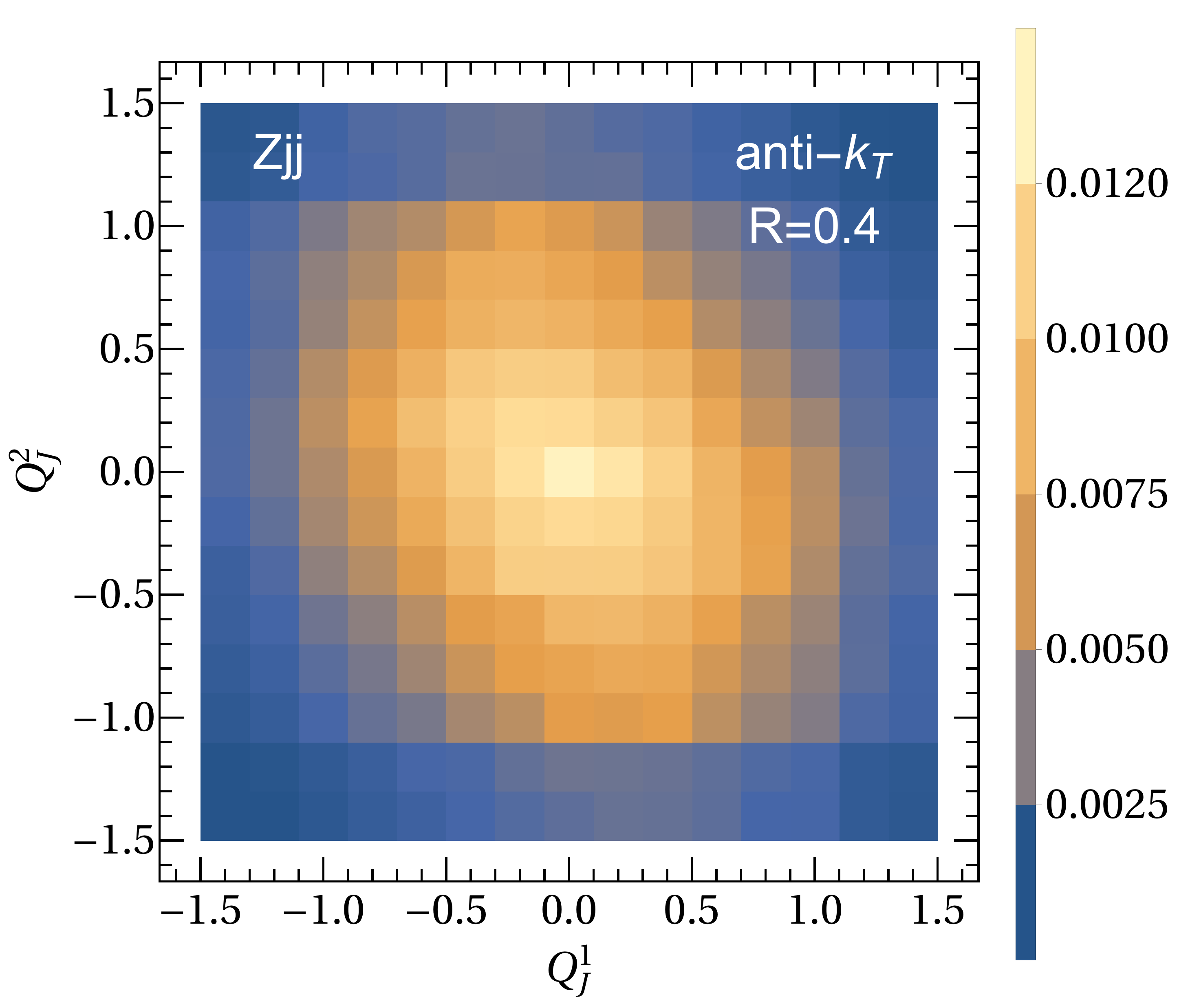}
\caption{The jet charge correlations between the leading and subleading jets from the signal and backgrounds with the jet charge parameter $\kappa=0.3$ under anti-$k_T$ algorithm with jet raidus $R=0.4$.}
\label{Fig:jQ}
\end{figure*}

\begin{figure*}
\centering
\includegraphics[scale=0.2]{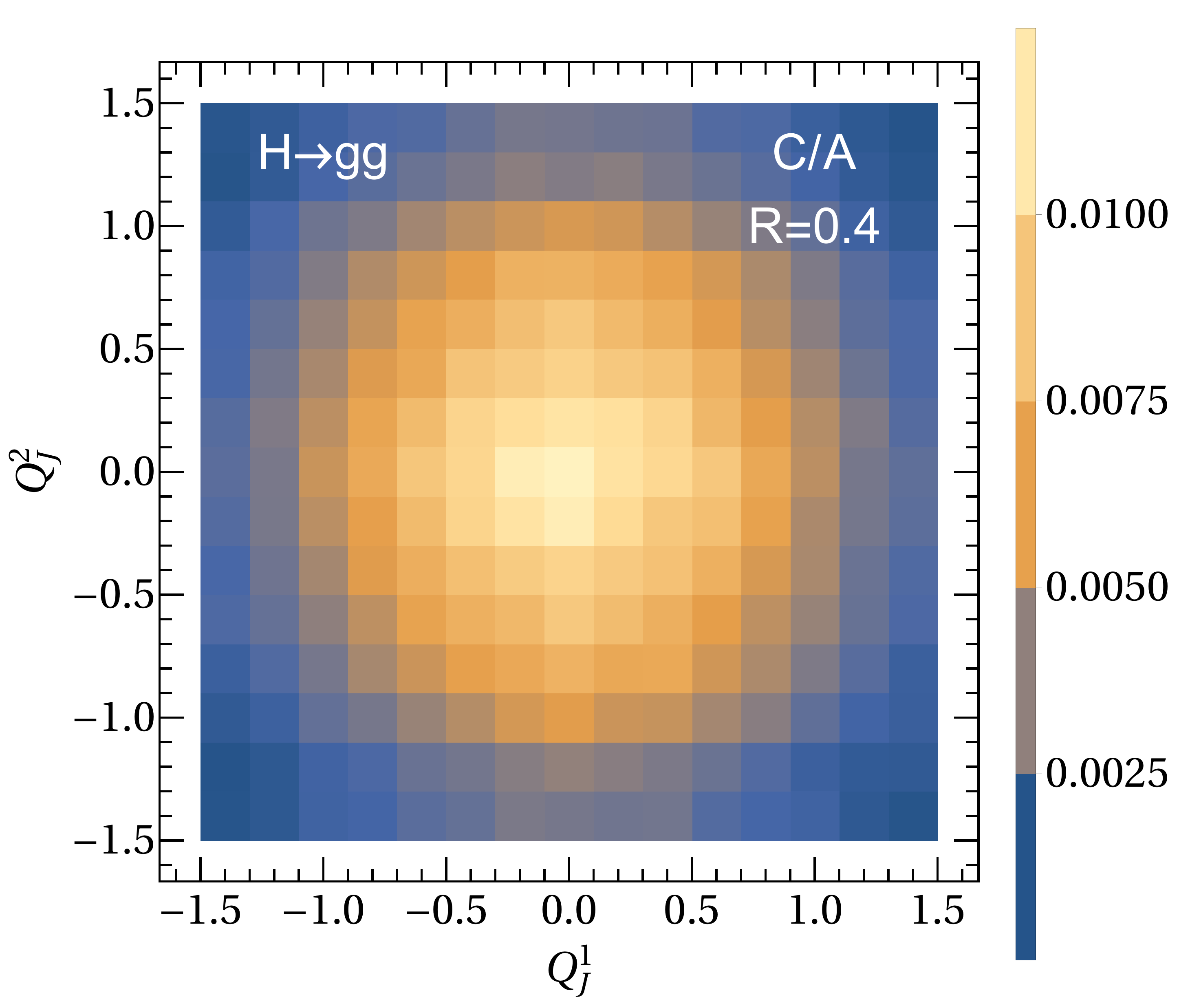}
\includegraphics[scale=0.2]{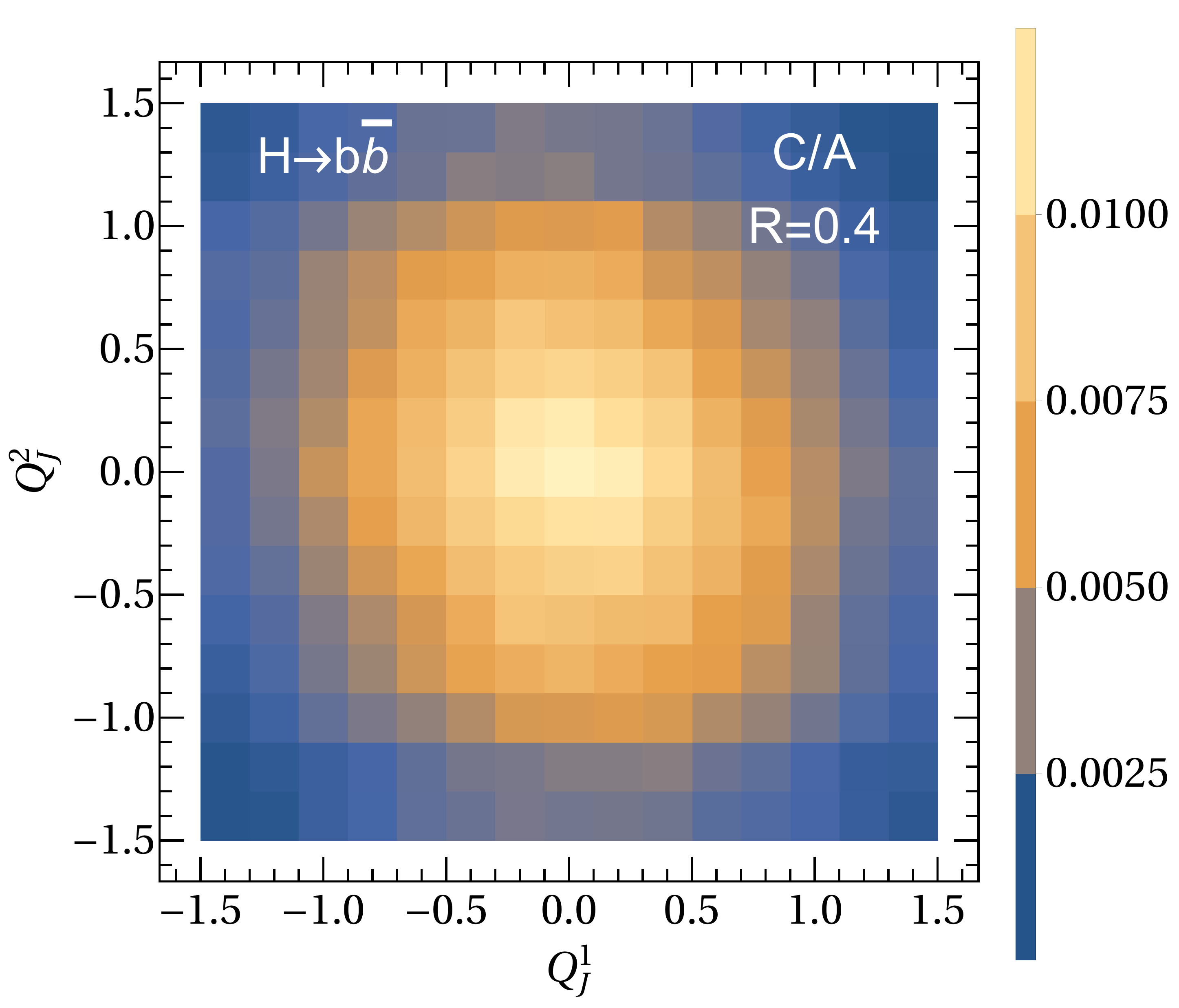}
\includegraphics[scale=0.2]{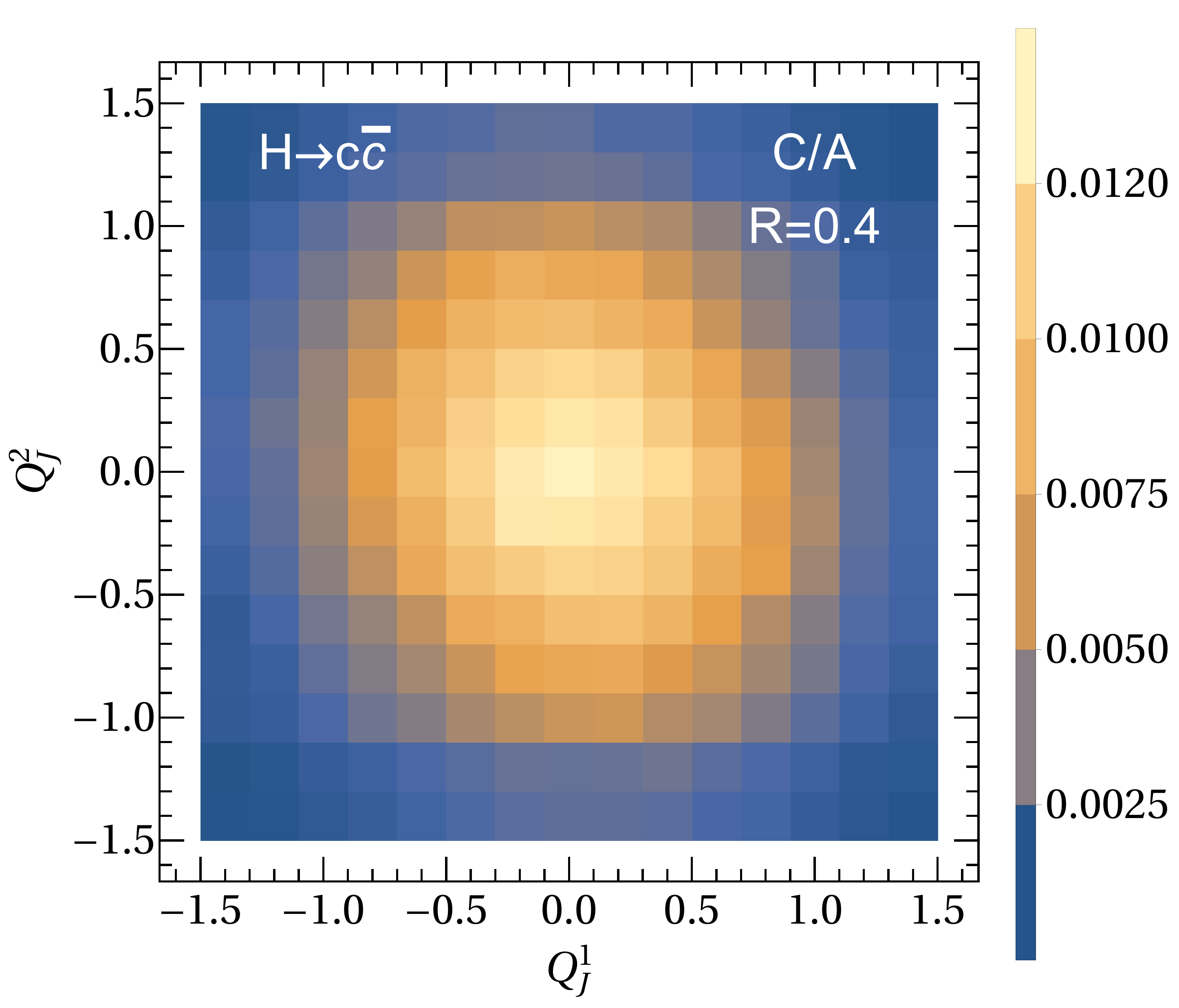}
\includegraphics[scale=0.2]{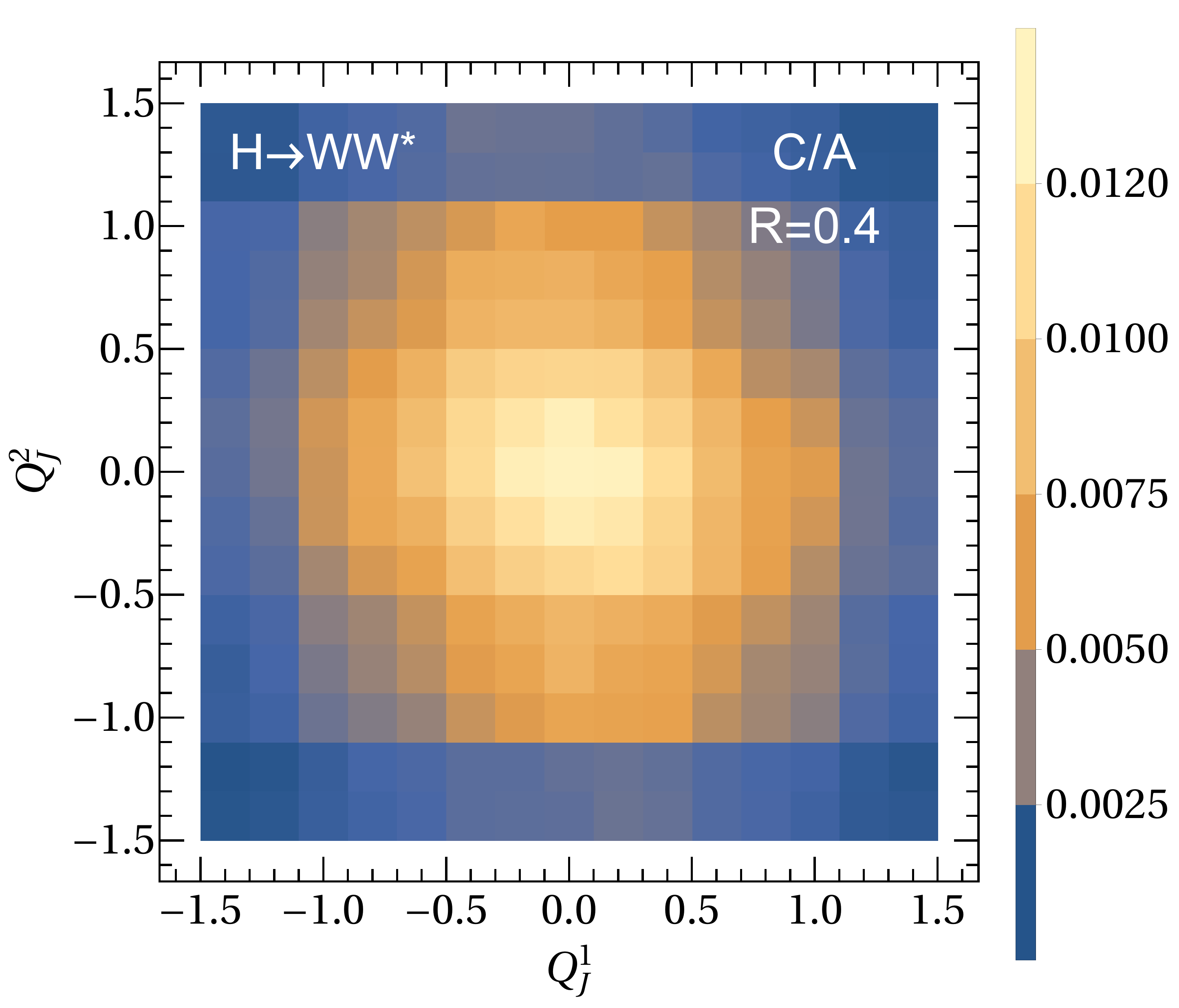}
\includegraphics[scale=0.2]{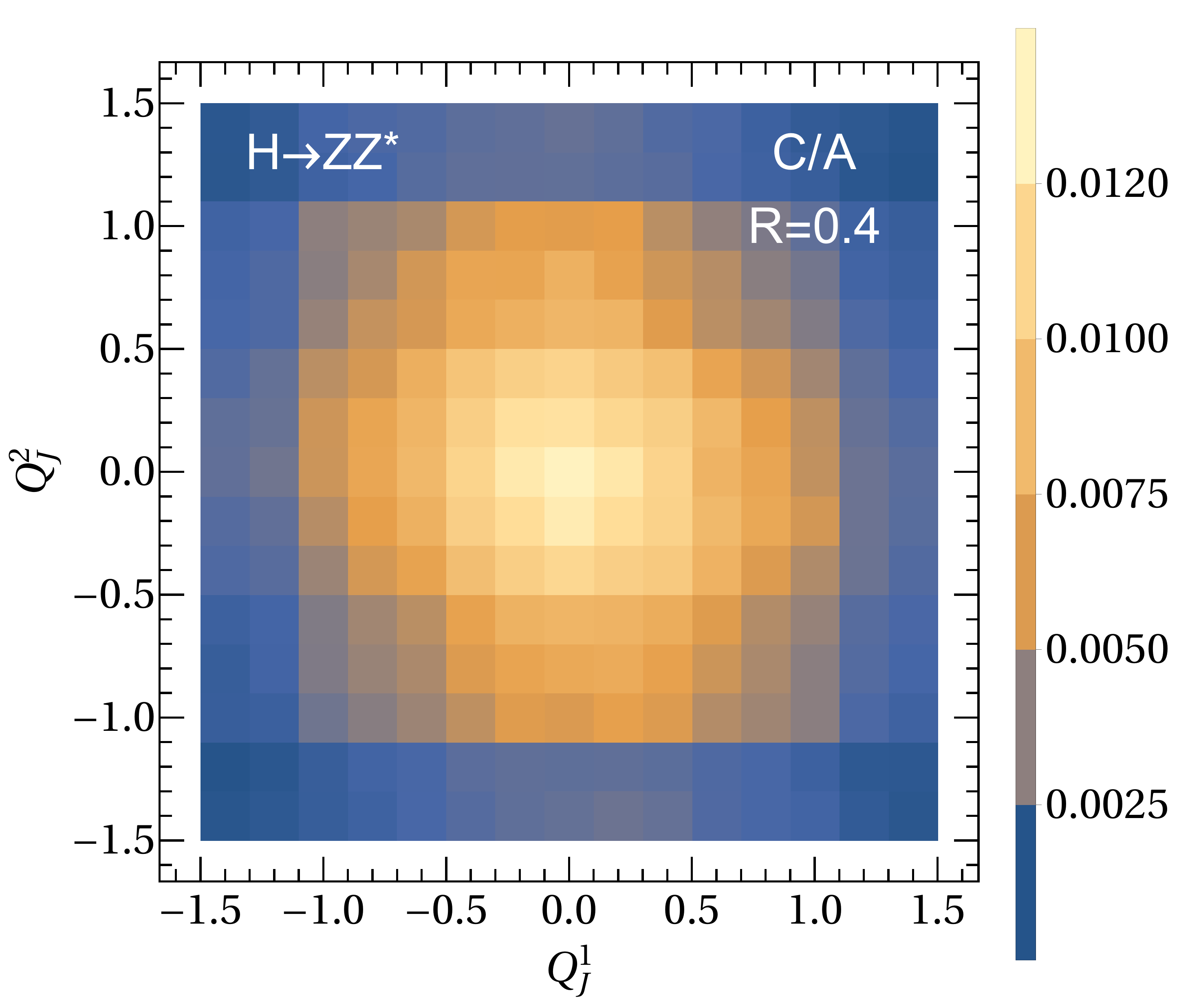}
\includegraphics[scale=0.2]{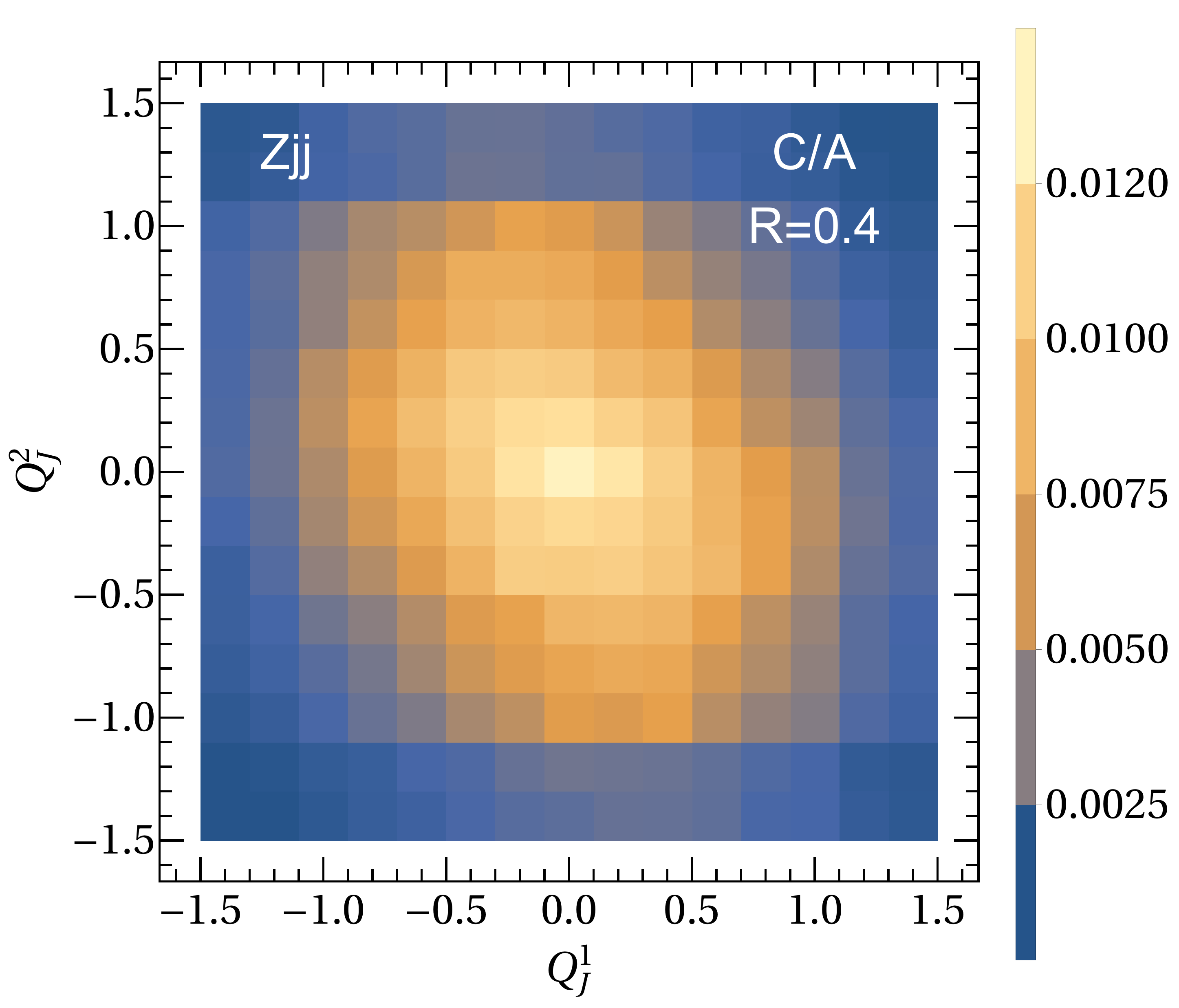}
\caption{Similar to Fig.~\ref{Fig:jQ}, but with Cambridge/Aachen(C/A) algorithm.}
\label{Fig:jQca}
\end{figure*}

\begin{figure*}
\centering
\includegraphics[scale=0.2]{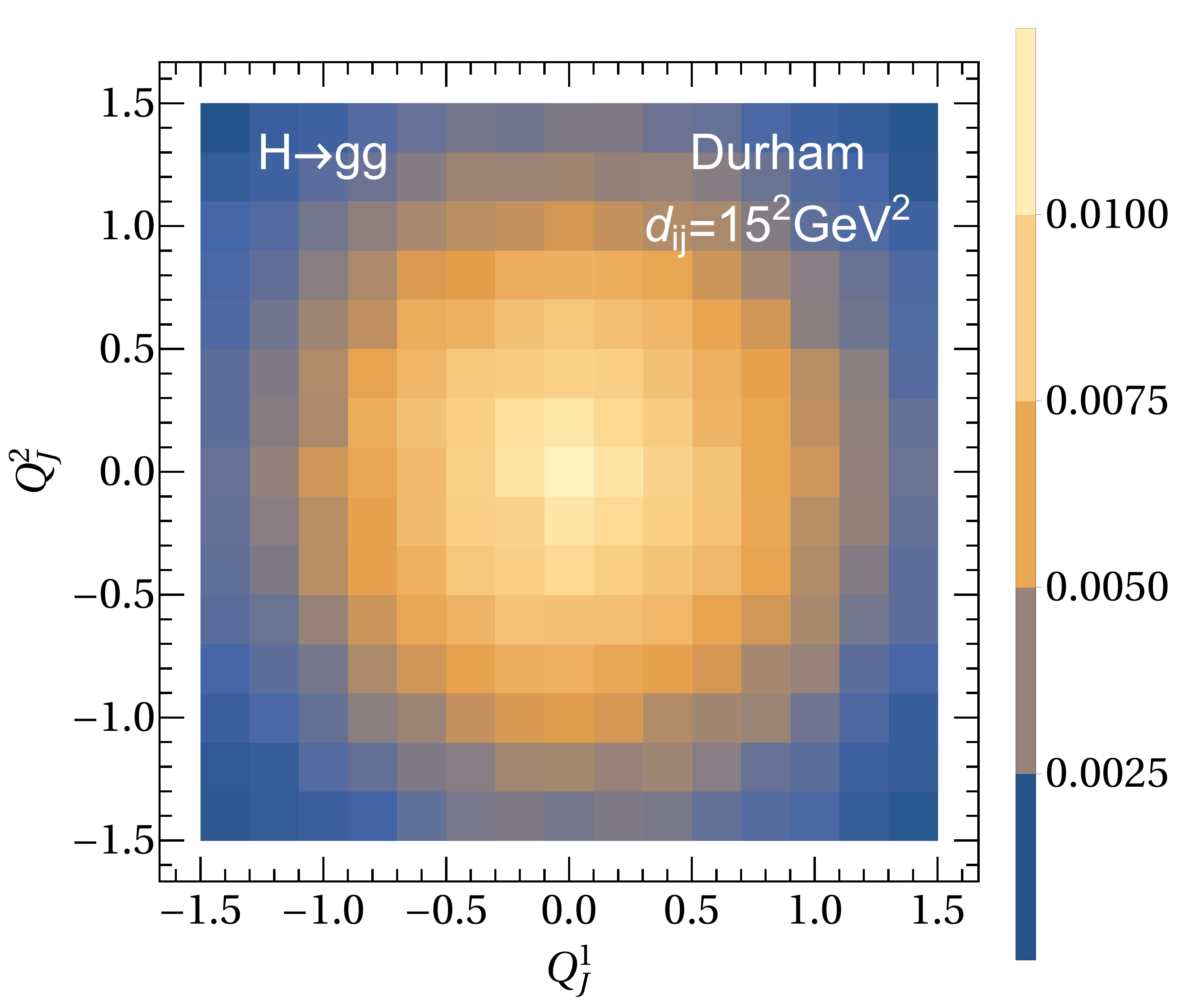}
\includegraphics[scale=0.2]{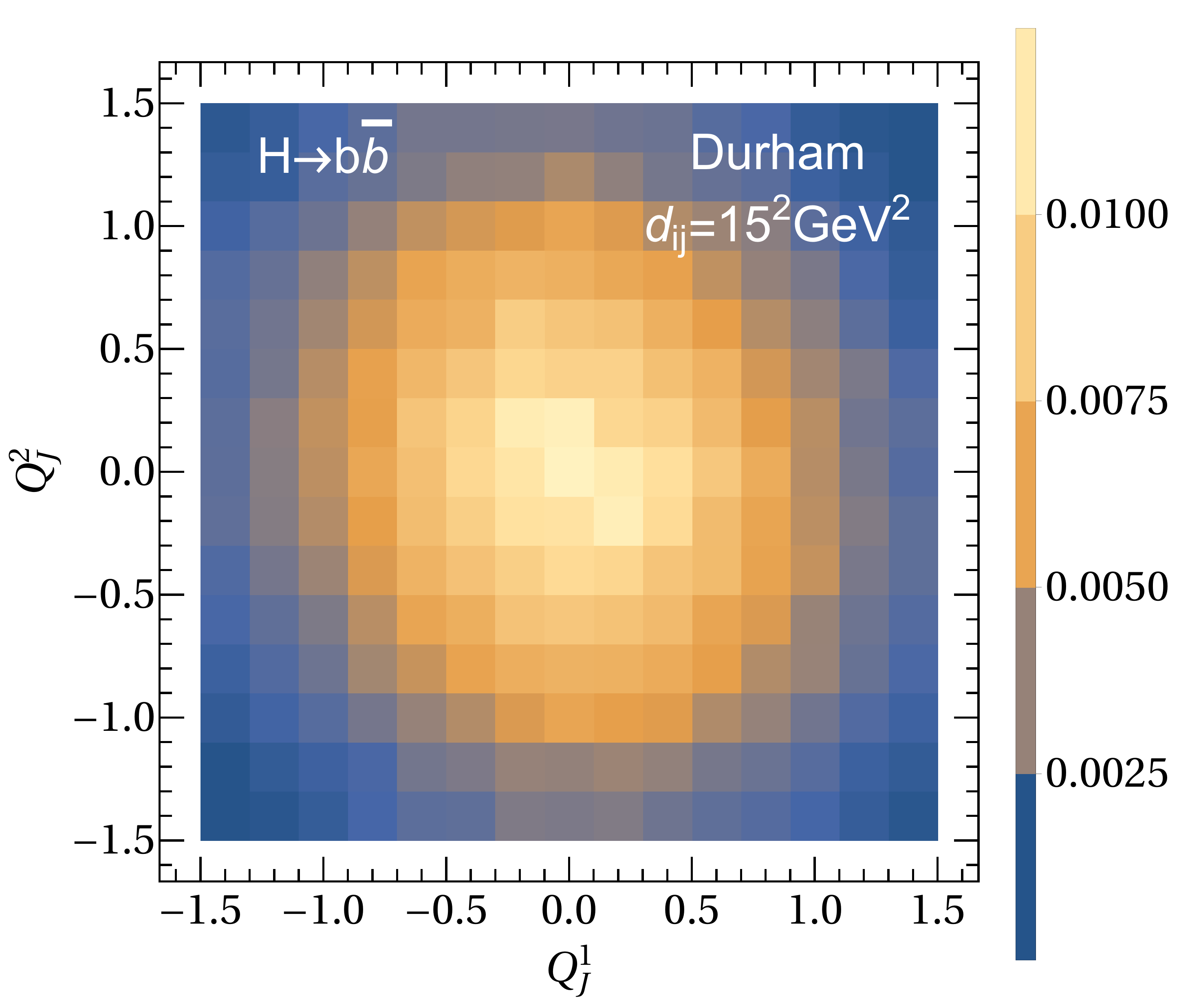}
\includegraphics[scale=0.2]{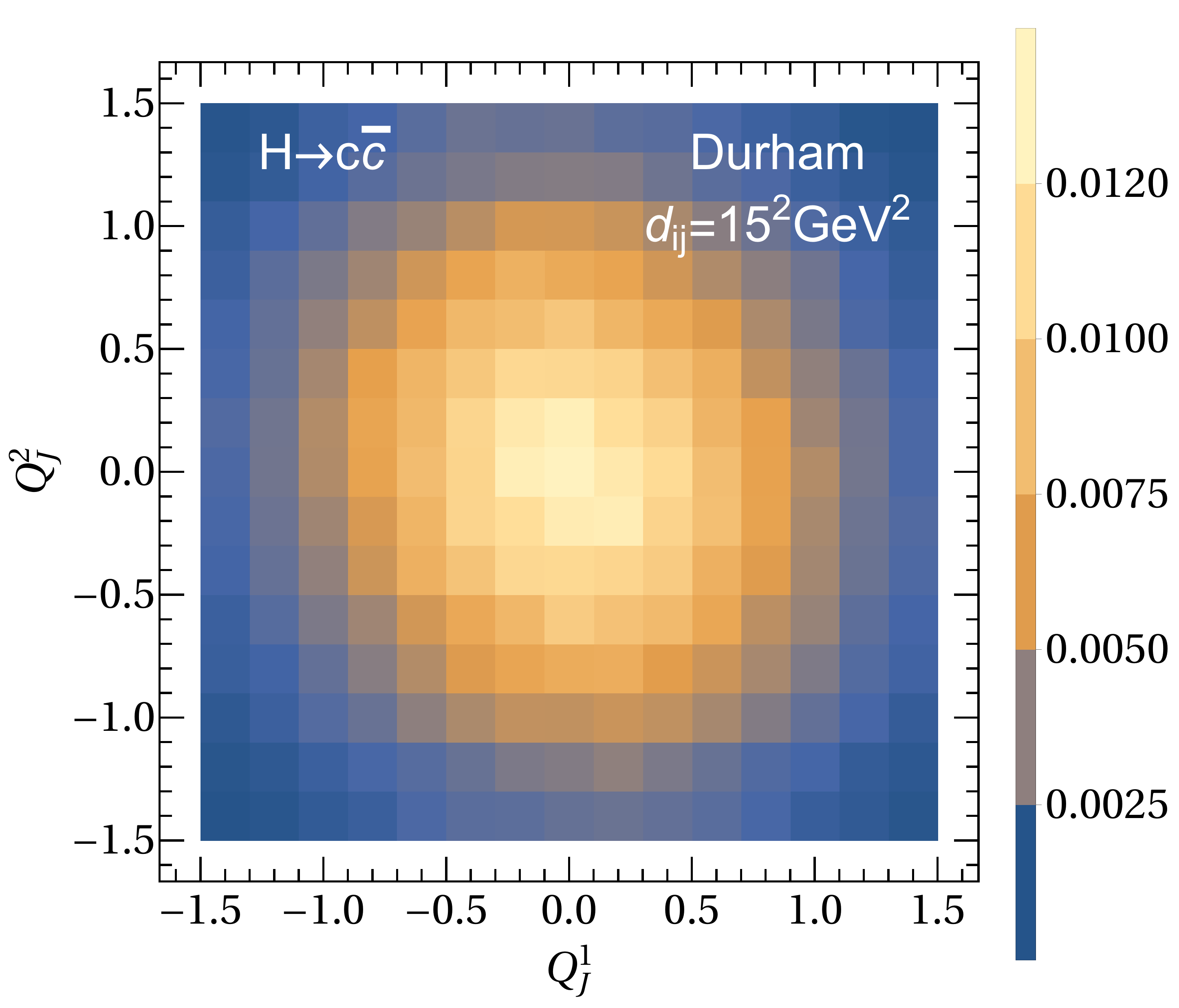}
\includegraphics[scale=0.2]{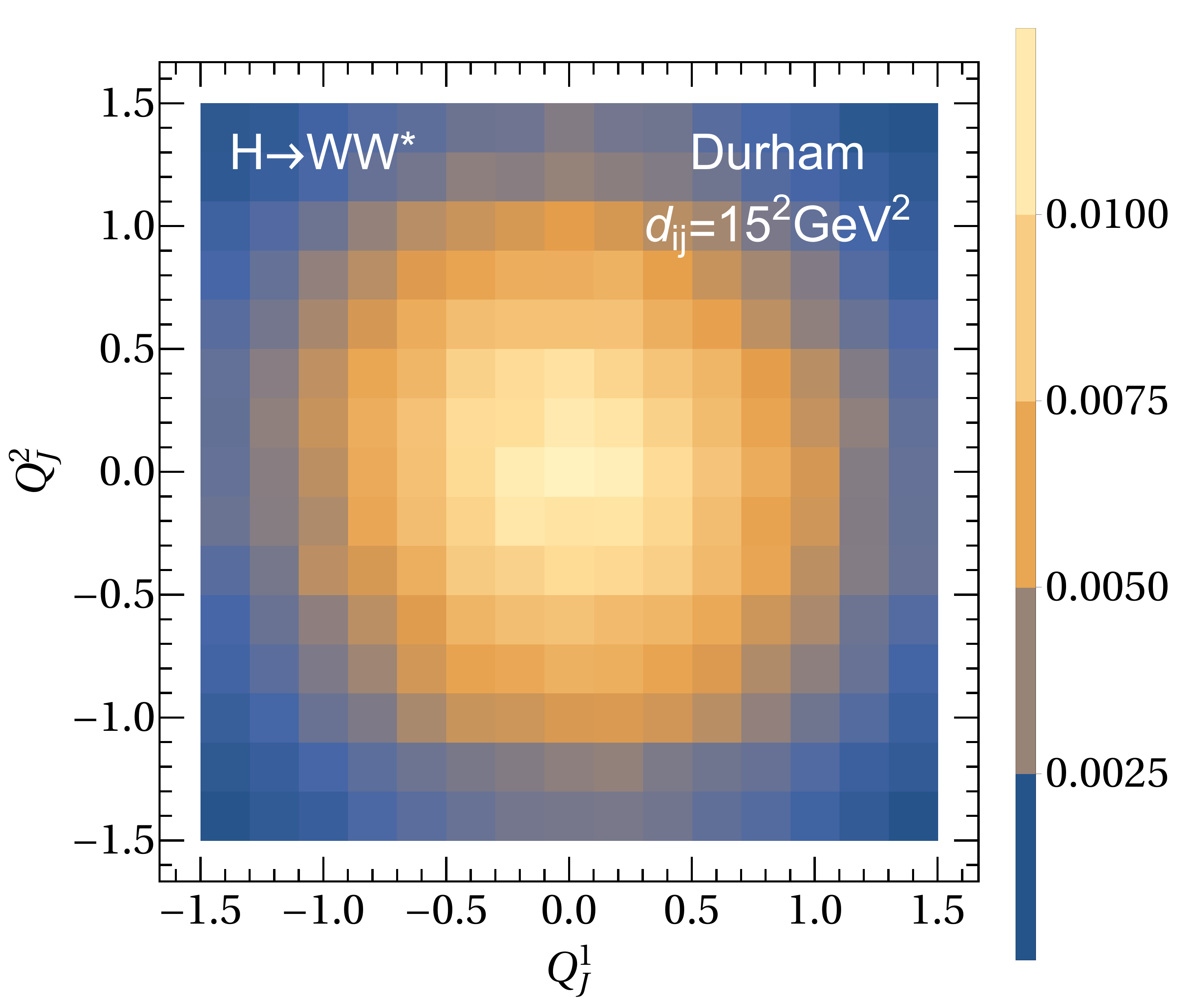}
\includegraphics[scale=0.2]{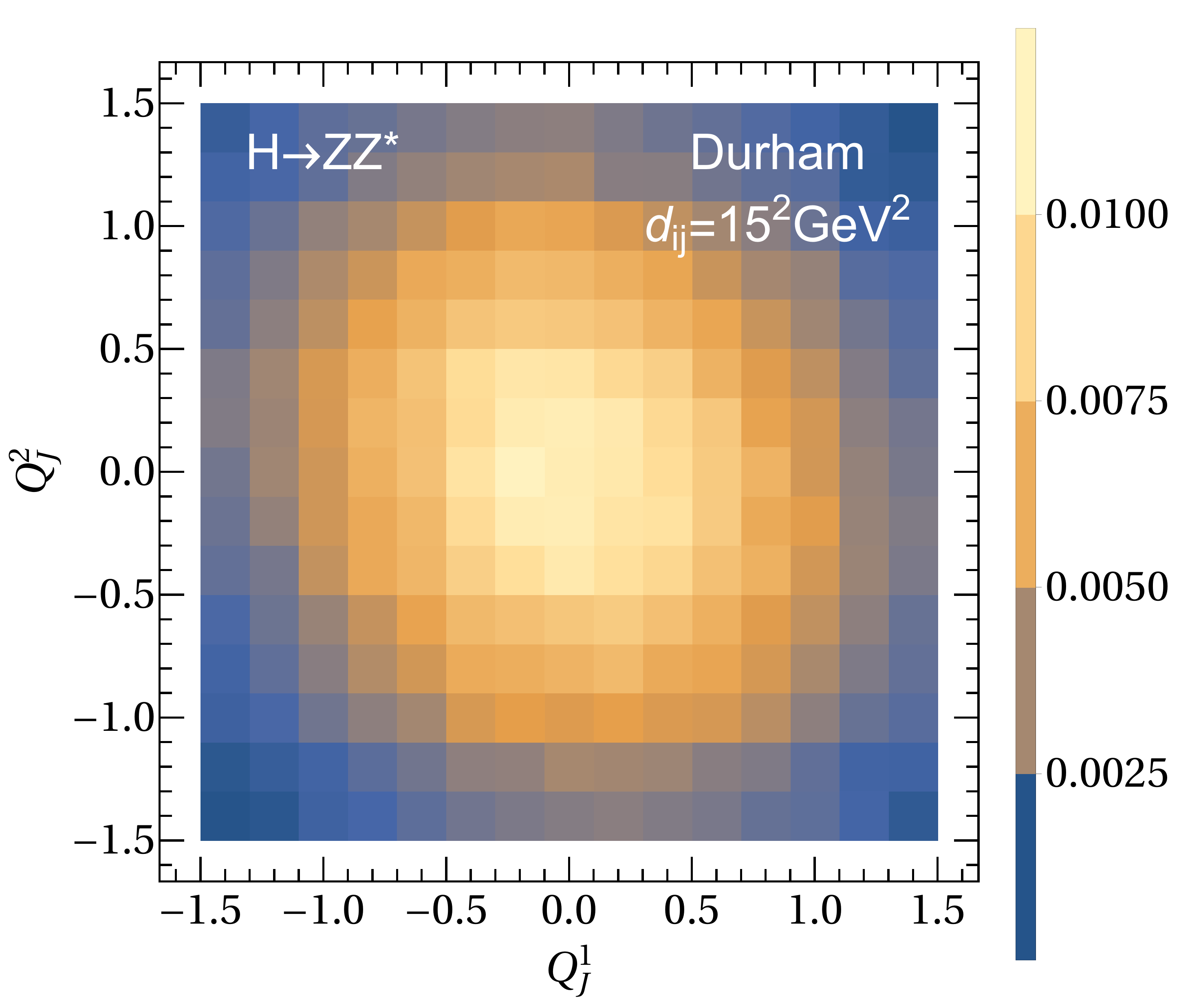}
\includegraphics[scale=0.2]{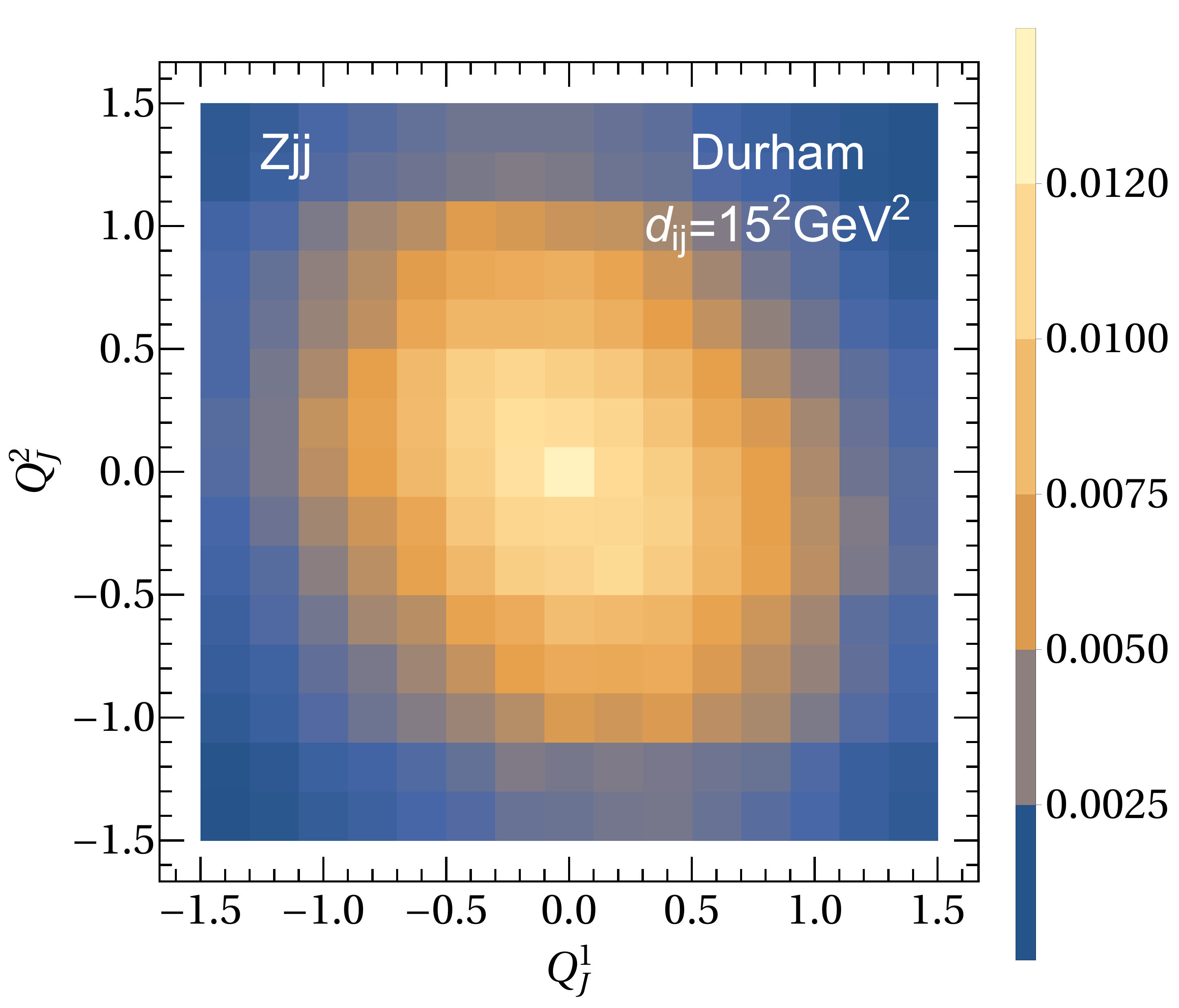}
\caption{Similar to Fig.~\ref{Fig:jQ}, but with Durham algorithm. Here $d_{ij}=15^2{\rm GeV}^2$, which is the distance parameter under the convention in Ref.~\cite{Cacciari:2011ma}.}
\label{Fig:jQdu}
\end{figure*}

Next, we utilize the event generator MadGraph5~\cite{Alwall:2014hca} to generate the signal and backgrounds events at the parton level and pass them to the PYTHIA~\cite{Sjostrand:2014zea,Bierlich:2022pfr}\footnote{It has been demonstrated in Fig. 1 of Ref.~\cite{ATLAS:2015rlw} and Figs. 2 and 3 of Ref.~\cite{CMS:2017yer} from the ATLAS and CMS Collaborations, the various hadronization models, such as PYTHIA and HERWIG, could give a similar jet charge distribution.} for parton showering and hadronization. We further require  $E_j>20~{\rm GeV}$ and $|\cos\theta|<0.99$, where $E_j$ and $\theta$ denotes the energy and polar angle of the jet, respectively. 

Figures \ref{Fig:jQ}-\ref{Fig:jQdu} show the jet charge correlations between the leading and subleading jets from the signal and background processes with the benchmark jet charge parameter $\kappa=0.3$ under anti-$k_T$~\cite{cacciari:2008gp}, Cambridge/Aachen(C/A)~\cite{Dokshitzer:1997in} and Durham~\cite{Catani:1991hj} jet clustering algorithms. As we expected that the jet charges are uncorrelated for the signal, but it exhibits a strong correlation for all the backgrounds. Additionally, we found that the jet charge correlations are not sensitive to the choice of jet clustering algorithms, therefore, we will use C/A algorithm as a benchmark in the following analysis.

\begin{figure}
\centering
\includegraphics[scale=0.35]{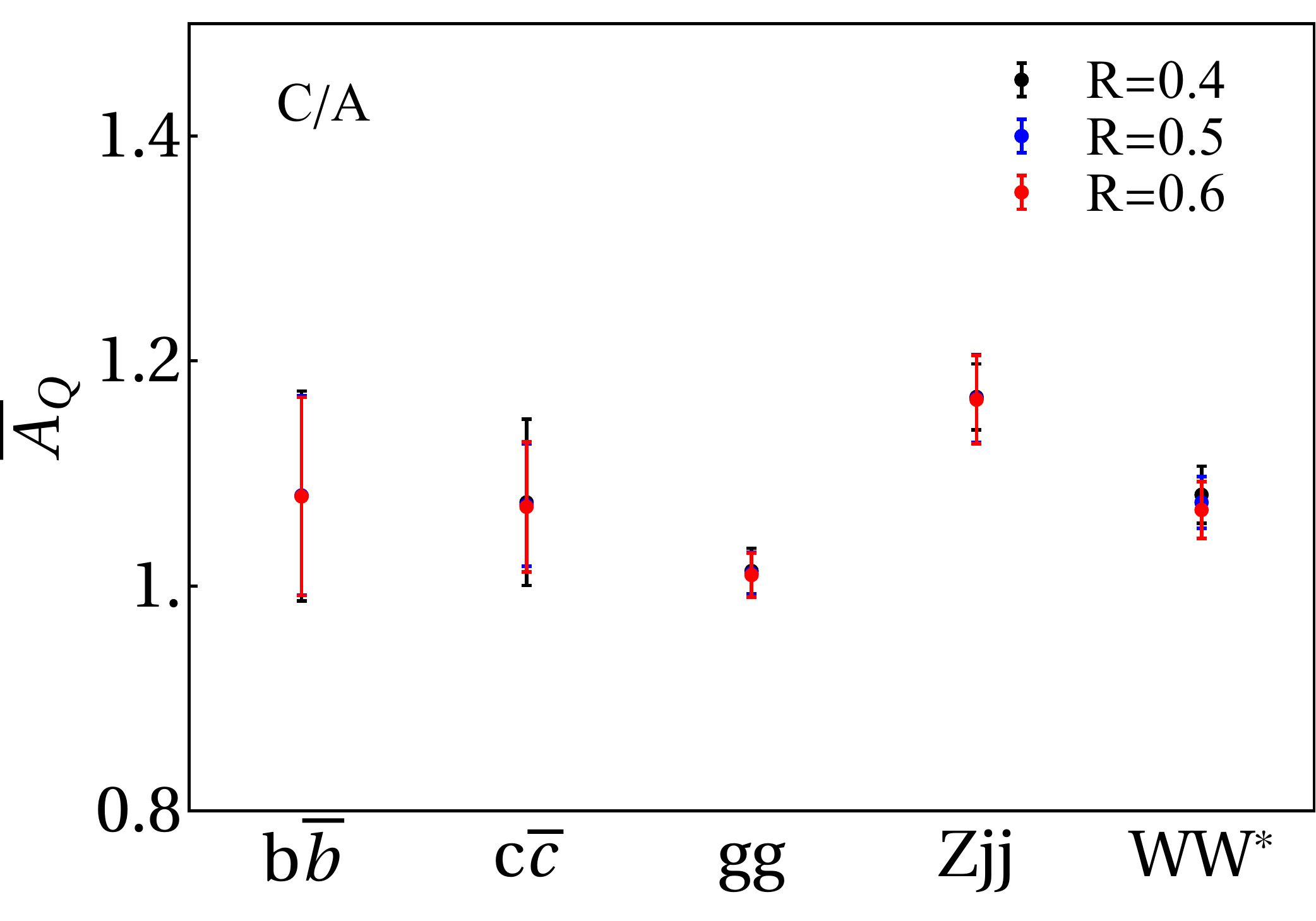}
\caption{The jet charge asymmetries of the signal and backgrounds under the C/A jet algorithm with $R=0.4,0.5,0.6$  at the CEPC with $\sqrt{s}=250~{\rm GeV}$ for the integrated luminosity $\mathcal{L}=5.6~{\rm ab}^{-1}$. The systematic uncertainties are assumed to be canceled from the definition.}
\label{Fig:AQ}
\end{figure}

\begin{figure}
\centering
\includegraphics[scale=0.42]{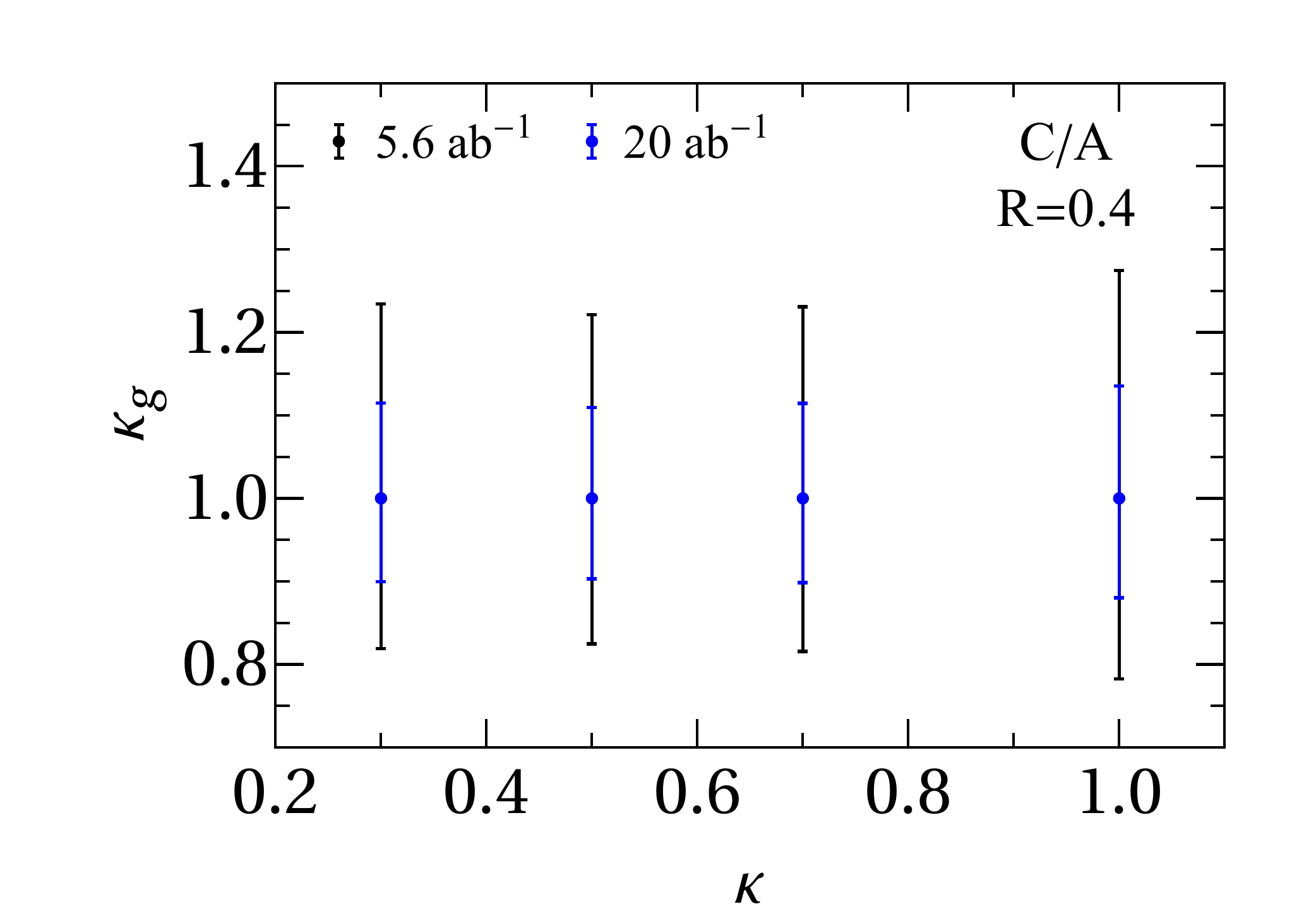}
\caption{The expected limits on the Higgs gluon effective coupling with the different jet charge parameter $\kappa$ under the C/A jet algorithm with $R=0.4$ at the CEPC with $\sqrt{s}=250~{\rm GeV}$ for the integrated luminosity $\mathcal{L}=5.6~{\rm ab}^{-1}$ (black points) and $\mathcal{L}=20~{\rm ab}^{-1}$ (blue points).}
\label{Fig:kg}
\end{figure}

Combing the jet charge information, we calculate the jet charge asymmetries for the signal and backgrounds under the C/A jet algorithm with cone size $R=0.4,0.5,0.6$ at the $\sqrt{s}=250~{\rm GeV}$ CEPC with the integrated luminosity $\mathcal{L}=5.6~{\rm ab}^{-1}$ . A large number of pseudo experiments have been generated by the PYTHIA for estimating the  statistical uncertainty of $\overline{A}_Q$, while the systematic errors are expected to be canceled from the jet charge asymmetry definition; see Eq.~\eqref{eq:Asy}. We assume that the statistical error satisfies the Gaussian distribution and could be rescaled based on the event number of the process. It clearly shows that $\overline{A}_Q\sim 1$ for the signal $H\to gg$, while $\overline{A}_Q>1$ for the backgrounds, as we argued before. Note that $\overline{A}_Q$ from the $H\to ZZ^*$ is similar to the $H\to WW^*$, but with a much larger statistical error, as a result, it is not plotted in Fig.~\ref{Fig:AQ}. It also shows that the $\bar{A}_Q$ is not sensitive to the jet radius, therefore, we adopt $R=0.4$ as a benchmark study in our analysis.

Now we combine all the jet charge information to constrain the $Hgg$ coupling. The combined jet charge asymmetry can be written as,
\begin{align}
\overline{A}_Q^{\rm tot}=
    & \frac{\sum_i f_i \langle Q^{(-)}\rangle_i }{\sum_i f_i \langle Q^{(+)}\rangle_i} \;,
\label{eq:Asy2}
\end{align}
where $f_i$ denotes the fraction of channel $i$, which was determined by the event numbers after the kinematic cuts. To estimate the event numbers from $H\to b\bar{b}$ and $H\to c\bar{c}$ modes, the mistagging efficiencies of $\epsilon_{b\to j}=8.9\%$ and $\epsilon_{c\to j}=40.7\%$ have been used~\cite{Gao:2016jcm}. We should note that the fractions $f_{gg},f_{VV}\gg f_{Zjj}$, thus, the asymmetry $\overline{A}_Q^{\rm tot}$ would be not sensitive to the Higgs production cross section and total width; see Eq.~\eqref{eq:Asy2}. To estimate the uncertainty of $\kappa_g$ from the $\overline{A}_Q^{\rm tot}$, we use the error propagation equation to calculate the error of $\overline{A}_Q^{\rm tot}$, {\it i.e.,}
\beq
\frac{\delta\overline{A}_Q^{\rm tot}}{\overline{A}_Q^{\rm tot}}=\sqrt{\left(\frac{\delta \langle Q^{(-)}\rangle}{\langle Q^{(-)}\rangle}\right)^2+\left(\frac{\delta \langle Q^{(+)}\rangle}{\langle Q^{(+)}\rangle}\right)^2},
\eeq
where $\delta \langle Q^{(\pm)}\rangle$ is the statistical error of numerator and denominator of  $\overline{A}_Q^{\rm tot}$, and
\beq
\delta \langle Q^{(\pm)}\rangle=\sqrt{\sum_i f_i^2\left(\delta \langle Q^{(\pm)}\rangle_i\right)^2}.
\eeq
Here $\delta \langle Q^{(\pm)}\rangle_i$ is the statistical error of $\langle Q^{(\pm)}\rangle_i$ of the $i$-th channel. Figure~\ref{Fig:kg} shows the uncertainties of the $Hgg$ coupling modifier $\kappa_g$ at the CEPC with different jet charge parameters at the 68\% confidence level. We note that the typical uncertainty of $\kappa_g$ from the jet charge asymmetry is around 20\% for the $\mathcal{L}=5.6~{\rm ab}^{-1}$ and the conclusion is not sensitive to the jet charge parameter $\kappa$. According to the updated note of CEPC group in Ref.~\cite{CEPCPhysicsStudyGroup:2022uwl}, the integrated luminosity could be upgraded to $20~{\rm ab}^{-1}$ and the error of $\kappa_g$ could be reduced to around 10\%. It is evident that the expected uncertainty of $\kappa_g$ from the jet charge asymmetry would be 3-4 times larger than the measurements from the jet energy profile~\cite{Li:2018qiy,Li:2019ufu} and the signal strength with $H\to gg$~\cite{An:2018dwb}. However, the different features of the jet charges from the signal and backgrounds also suggest that it could be complementary to other methods for determining $\kappa_g$ when we use advanced technologies such as Boosted Decision Tree (BDT) or Machine Learning in experimental analysis, which is however beyond the scope of this work.

\section{Conclusions}
In this paper, we propose to constrain the $Hgg$ coupling via the jet charge asymmetry $\overline{A}_Q$ from the Higgs boson decay at the future lepton collider. Owing to the jet charges being fully uncorrelated for the signal $H\to gg$, we demonstrated that $\overline{A}_Q\sim 1$, while the charges from the dominated backgrounds are correlated, as a result, we obtain $\overline{A}_Q>1$. Such a different feature can be used to separate the $H\to gg$ from the backgrounds and to constrain the $Hgg$ coupling. It shows that the uncertainty of the coupling modifier $\kappa_g$ from the jet charge information would be much larger than the traditional methods at the CEPC with an integrated luminosity $\mathcal{L}=5.6~{\rm ab}^{-1}$. But we emphasize that the jet charge information is still useful for determining the $\kappa_g$ if we can combine the jet charge information and other kinematic distributions with the BDT or Machine Learning in the analysis.

\vspace{3mm}
\medskip
\noindent{\bf Acknowledgments.}
The authors thank Rui Zhang for the helpful discussion. This work is supported by the IHEP under Contract No. E25153U1.

\bibliographystyle{apsrev}
\bibliography{reference}

\end{document}